\documentclass[12pt,letterpaper]{article}
\usepackage[a4paper, total={7in, 10in}]{geometry}

\usepackage{multirow}
\usepackage{graphicx}
\usepackage{geometry}
\usepackage{bm}
\usepackage{helvet}
\usepackage{xspace}
\usepackage{authblk}
\usepackage{hyperref}
\usepackage{amsmath} 
\usepackage{amssymb} 
\usepackage{orcidlink} 
\usepackage[super,comma,sort&compress]  
   {natbib}
\usepackage[right]{lineno} \linenumbers

\newcommand{\CrO}{$\mathrm{Cr}_{2}\mathrm{O}_{3}$}

\makeatletter
\renewcommand{\maketitle}{\bgroup\setlength{\parindent}{0pt}
\begin{flushleft}
  \textbf{\@title}
  
  \@author
\end{flushleft}\egroup}
\makeatother

\title{THz electric field control of spins in collinear antiferromagnet Cr$_{2}$O$_{3}$}
\date{}

\author[1,*]{V.~R.~Bilyk\,\orcidlink{0000-0002-3013-8655}}
\author[2]{R.~M.~Dubrovin\,\orcidlink{0000-0002-7235-7805}}
\author[3,4]{A.~K.~Zvezdin\,\orcidlink{0000-0002-6039-780X}}
\author[1]{A.~I.~Kirilyuk\,\orcidlink{0000-0003-1479-9872}}
\author[1]{A.~V.~Kimel\,\orcidlink{0000-0002-0709-042X}}
\affil[1]{Institute for Molecules and Materials, Radboud University, 6525 AJ Nijmegen, The Netherlands}
\affil[2]{Ioffe Institute, Russian Academy of Sciences, 194021 St.\,Petersburg, Russia}
\affil[3]{New Spintronic Technologies LLC, 121205 Skolkovo, Moscow, Russia}
\affil[4]{Prokhorov General Physics Institute, Russian Academy of Sciences, 119991 Moscow, Russia}
\affil[*]{Correspondence: \href{mailto:vladislav.bilyk@ru.nl}{vladislav.bilyk@ru.nl}}
\date{} 

\begin{document}

\maketitle

\section*{SUMMARY}

The idea to find a magnet that responds to an electric field as efficiently as to its magnetic counterpart has long intrigued people's minds and recently became a cornerstone for future energy efficient and nano-scalable technologies for magnetic writing and information processing~\cite{yang2023terahertz,mostovoy2024multiferroics,hassanpour2022magnetoelectric,leo2018magnetoelectric,matsukura2015control,lottermoser2004magnetic}.
In contrast to electric currents, a control by electric fields promises much lower dissipations and in contrast to magnetic fields, electric fields are easier to apply to a nanoscale bit.
Recently, the idea to find materials and mechanisms facilitating a strong and simultaneously fast response of spins to electric field has fueled an intense research interest to electromagnons in non-collinear antiferromagnets~\cite{mostovoy2024multiferroics,yang2023terahertz,wu2023fluctuation,verseils2023stabilizing,masuda2021electric,toth2016electromagnon,jones2014high,kubacka2014large,pimenov2006possible,gao2024giant}.
Here we show that THz spin resonance at the frequency 0.165\,THz in collinear antiferromagnet Cr$_{2}$O$_{3}$, which does not host any electromagnons, can be excited by both THz magnetic and electric fields.
The mechanisms result in comparable effects on spin dynamics, when excited by freely propagating electromagnetic wave, but have different dependencies on the orientation of the applied THz electric field and the antiferromagnetic N{\'e}el vector.
Hence this discovery opens up new chapters in the research areas targeting to reveal novel principles for the fastest and energy efficient information processing - ultrafast magnetism, antiferromagnetic spintronics, and THz magnonics.

\section*{KEYWORDS}

Antiferromagnet, magnetoelectric, magnon, spin dynamics,  terahertz excitation 

\section*{INTRODUCTION}

Current technologies are primarily based on the most well-known class of ferromagnetic materials.
In ferromagnets, electron spins are aligned mutually parallel, resulting in a net magnetization $\mathbf{M}$.
Due to this magnetization, spins of ferromagnets can be switched between stable bit states by reversing the magnetization by the magnetic field or electrical currents, facilitating writing at rates in the order of 1\,GHz.
However, nature exhibits another, and, in fact, much more abundant class of magnetic materials called antiferromagnets, where the spins are mutually antiparallel.
In the simplest case, an antiferromagnet can be considered as two coupled ferromagnets with mutually antiparallel net magnetizations $\mathbf{M}_{\mathrm{A}}$ and $\mathbf{M}_{\mathrm{B}}$ ($\mathbf{M}_{\mathrm{A}}=-\mathbf{M}_{\mathrm{B}}$), respectively.
While $\mathbf{M} = \mathbf{M}_{\mathrm{A}} + \mathbf{M}_{\mathrm{B}} = 0$, the spin order in antiferromagnets is described by the antiferromagnetic N{\'e}el vector $\mathbf{L} = \mathbf{M}_{\mathrm{A}} - \mathbf{M}_{\mathrm{B}} \neq 0$.
Similarly to ferromagnets, a reversal of the antiferromagnetic N{\'e}el vector switches an antiferromagnet between stable bit states, but, in theory, the switching in antiferromagnets can be up to 1000 times faster than their ferromagnetic counterparts~\cite{han2023coherent,nemec2018antiferromagnetic,baltz2018antiferromagnetic,jungwirth2016antiferromagnetic}. 
Moreover, distinct stable states of an antiferromagnet have equal energies, equal entropies, and, contrary to ferromagnets, equal angular momenta.
Hence, writing antiferromagnetic bits does not imply irreversible transfer of energy or angular momentum from the spins.
Therefore, antiferromagnets represent an intriguing playground to search not only for the fastest, but also for the least dissipative mechanism of data storage.
Understanding the mechanisms and scenarios to control the antiferromagnetic N{\'e}el vector $\mathbf{L}$ has been seen as a challenge from the very discovery of antiferromagnetism and remains to be a hot topic until now ~\cite{Qiu2023Axion, Du2023Antiferromagnetic}. 

Indeed while the magnetization $\mathbf{M}$ is the thermodynamic conjugate to the magnetic field $\mathbf{H}$, the antiferromagnetic N{\'e}el vector is not i.e  $\mathbf{L} \cdot \mathbf{H} = 0$ suggesting that the field is unable to create a torque on $\mathbf{L}$.
THz magnetic field was shown to be a game changer in the field~\cite{satoh2010spin,kampfrath2011coherent}.
It efficiently launches spin oscillations in collinear antiferromagnets, when applied perpendicular to the antiferromagnetic N{\'e}el vector and such that its time-derivative $\dot{\mathbf{H}} \perp \mathbf{L}$ plays the role of the driving force~\cite{zvezdin1979dynamics,zvezdin2017dynamics,mashkovich2021terahertz}. 
Also THz electric current are claimed to control spins in antiferromagnets~\cite{Behovits2023terahertz}.
The discovery of the effect of electric field on spins in antiferromagnetic \CrO{} was practically the very first experimental demonstration of magnetoelectric effect~\cite{astrov1960magnetoelectric,astrov1961magnetoelectric}.
It is thus interesting to verify if THz electric field can excite spin resonance in the collinear antiferromagnet \CrO{}, if the mechanism is strong and how it is different from the well understood control of spins in antiferromagnets by the THz magnetic field. 

\section*{RESULTS AND DISCUSSION}

\begin{figure*}[ht]
\centering
\includegraphics[width=1\columnwidth]{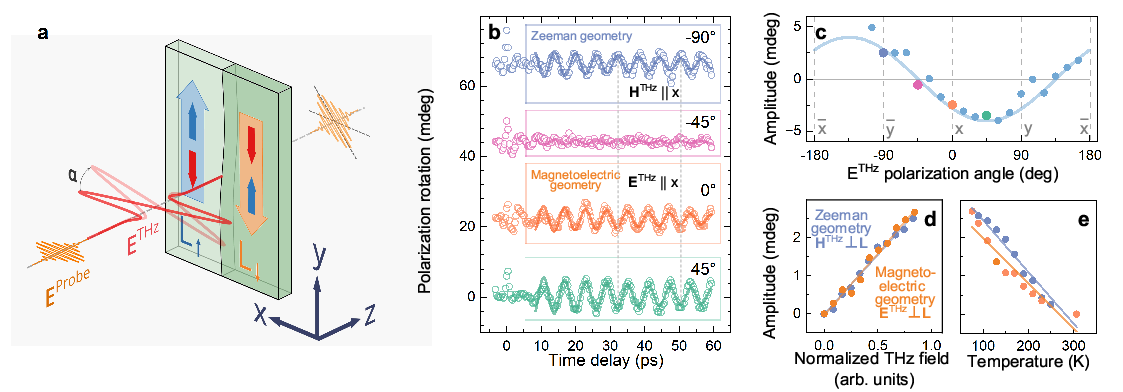}
\caption{\label{fig:1}
\textbf~The single-cycle THz pump pulse excites spin dynamics in magnetoelectric antiferromagnet \CrO{}. The dynamics is detected by measuring polarization rotation of the time-delayed linearly polarized infrared probe pulse. The polarization angle $\alpha$ and the intensity of the linearly polarized THz pump pulse were controlled with a set of two wire-grid polarizers.
The sample has two types of antiferromagnetic domains with mutually opposite antiferromagnetic N{\'e}el vectors $\mathbf{L}_{\uparrow}$ and $\mathbf{L}_{\downarrow}$ directed along the $y$ axis in the chosen Cartesian coordinates (crystallographic $c$ axis) .
\textbf{b}~Transient probe polarization rotation induced by THz electromagnetic pulse with different orientations of the THz electric field $E^{\mathrm{THz}}$ with respect to the $x$ axis. The measurements were done at 77\,K.
Circles are experimental data.
Lines are guides for the eye. 
\textbf{c}~The amplitude of excited oscillations as function of $\alpha$ -  the angle between the THz electric field $E^{\mathrm{THz}}$ with respect to the $x$ axis. Circles are experimental data.
Lines are fits as described in the text.
\textbf{d}~Dependence of the amplitude of oscillations on the THz electric $E^{\mathrm{THz}}$ and magnetic $H^{\mathrm{THz}}$ field strengths for the magnetoelectric ($\mathbf{E}^{\mathrm{THz}} \perp \mathbf{L}$, $\mathbf{H}^{\mathrm{THz}} \parallel \mathbf{L}$) and the Zeeman ($\mathbf{H}^{\mathrm{THz}} \perp \mathbf{L}$, $\mathbf{E}^{\mathrm{THz}} \parallel \mathbf{L}$) experimental geometries, respectively. 
\textbf{e}~Temperature dependence of the amplitude of the oscillations for the two geometries as deduced from the Fourier spectra of the experimentally observed time dependencies.
}
\end{figure*}

To answer these questions, we carried out pump-probe experiments in the geometry shown in Fig.~\ref{fig:1}a.
A single crystal of \CrO{} with the N{\'e}el temperature $T_{N} = 307$\,K and the antiferromagnetic N{\'e}el vector $\mathbf{L}$ aligned in the sample plane was excited by a nearly single-cycle THz electromagnetic pulse.
For convenience, we used a laboratory Cartesian coordinate system where the antiferromagnetic N{\'e}el vector $\mathbf{L}$ is along the $y$ axis and THz pump pulse propagates along the $z$ axis with the fields  $\mathbf{E}^{\mathrm{THz}}$ and $\mathbf{H}^{\mathrm{THz}}$ in the $xy$ plane. If the THz pulses launch spin dynamics, the latter must induce in the material an optical anisotropy, which was detected by measuring polarization rotation of the probe beam, as explained in detail in Methods.  

\begin{figure}[ht]
\centering
\includegraphics[width=1\columnwidth]{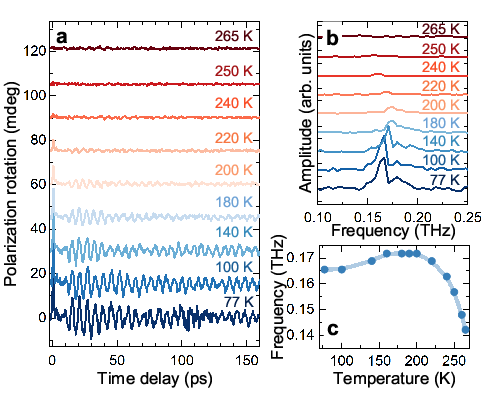}
\caption{\label{fig:S2}
\textbf{a}~THz pump induced probe polarization rotation measured at different temperatures in the Zeeman geometry $\mathbf{H}^{\mathrm{THz}} \perp \mathbf{L}$ ($\mathbf{E}^{\mathrm{THz}} \parallel \mathbf{L}$) in a single antiferromagnetic domain.
\textbf{b}~Fourier spectra of the waveforms from \textbf{a}.
\textbf{c} Temperature dependence of the magnon frequency.
}
\end{figure}

We observed the THz-induced transient signals for probe polarizations in \CrO{} below $T_{N}$ (Supplementary Note).
The Fourier spectra of these time traces reveal a resonance at 0.165\,THz at 77\,K.
The frequency of resonance and its temperature dependence (Supplementary Figure \ref{fig:S2}) are in fair agreement with those expected for antiferromagnetic resonance in \CrO{}~\cite{foner1963high,alikhanov1969neutron,samuelsen1970inelastic,biao2023low,qiyang2024temperature}.
The observed oscillations can thus be reliably assigned  the THz driven spin dynamics in the antiferromagnet.

According to the selection rules, efficient excitation of the antiferromagnetic resonance by the magnetic field $\mathbf{H}^{\mathrm{THz}}$ of the THz pump pulse (THz Zeeman torque) is expected, when the magnetic field is orthogonal to the antiferromagnetic N{\'e}el vector $\mathbf{H}^{\mathrm{THz}} \perp \mathbf{L}$~\cite{kampfrath2011coherent,baierl2016nonlinear}. 
On the contrary, when the magnetic field of the THz pump pulse is parallel to the antiferromagnetic N{\'e}el vector $\mathbf{H}^{\mathrm{THz}} \parallel \mathbf{L}$, and consequently $\mathbf{E}^{\mathrm{THz}} \perp \mathbf{L}$, the efficiency of this excitation mechanism must be equal to zero.

It is seen in Fig.~\ref{fig:1}b that the spin dynamics in \CrO{} is excited not only in the geometry, where the Zeeman torque is at maximum ($\mathbf{H}^{\mathrm{THz}} \perp \mathbf{L}$).  The excitation appears to be similarly efficient even if the Zeeman torque must be zero, where $\mathbf{H}^{\mathrm{THz}} \parallel \mathbf{L}$ ($\mathbf{E}^{\mathrm{THz}} \perp \mathbf{L}$). 
The amplitudes of oscillations, nevertheless, scale linearly with either of the THz fields ($\mathrm{E}^{\mathrm{THz}}$ or $\mathrm{H}^{\mathrm{THz}}$) as shown in Fig.~\ref{fig:1}d. The same finding is additionally confirmed by a phase shift of the observed oscillations by $\pi$, if the THz fields are rotated by $\pi$ in the $xy$ plane (see Supplementary Figure).

Figure~\ref{fig:1}c shows the amplitude of the oscillations as function of the angle $\alpha$ between the THz electric field $\mathbf{E}^{\mathrm{THz}}$ and the $x$ axis, where 
the sign change corresponds to the $\pi$ shift of phase of oscillations. The amplitudes were deduced from the Fourier spectra of the corresponding time dependencies (see Supplementary). Interestingly,  the amplitude is the largest at $\alpha=45^\circ$ and nearly zero  at $\alpha=-45^\circ$ as shown, in Figs.~\ref{fig:1}b and~\ref{fig:1}c. All these findings show that next to the conventional mechanism of excitation of spins in antiferromagnetic \CrO{} by the rapidly varying magnetic field $\mathbf{H}^{\mathrm{THz}}$, there is yet another, nearly as strong and sufficiently fast mechanism to excite the spins. Similarly to the Zeeman torque, however, the interfering mechanism must also be linear either with respect to $H^{\mathrm{THz}}$ or $E^{\mathrm{THz}}$.  
 
In order to obtain deeper insights into the origin of this alternative mechanism, we have developed a model of the THz pump driven spin dynamics in the magnetoelectric antiferromanget \CrO{}. The antiferromagnet is modelled in the two-sublattice approximation, where $\mathbf{m} = \mathbf{M} / |\mathbf{M}|$ and $\mathbf{l} = \mathbf{L} / |\mathbf{L}|$, as detailed in Methods.
The energy of interaction of spins with the electromagnetic fields in  \CrO{} may have three different contributions, which are linear either with respect to  $\mathbf{H}^{\mathrm{THz}}$ or $\mathbf{E}^{\mathrm{THz}}$ and allowed by the symmetry of the magnetoelectric antiferromagnet
\cite{belov1993magnetoelectric}
\begin{equation}
    \label{eq:U}
   U = - \lambda_{\perp} m_{x} l_{y} E^{\mathrm{THz}}_{x} - \lambda_{\parallel} m_{y} l_{y} E^{\mathrm{THz}}_{y} - \gamma S \hbar \left( m_{x} H^{\mathrm{THz}}_{x} + m_{y} H^{\mathrm{THz}}_{y} \right),
\end{equation}
where $\lambda$ is the parameter related to the magnetoelectric coefficient $\alpha$ from Ref.~\cite{astrov1960magnetoelectric}, $\gamma$ is the gyromagnetic ratio, and $S$ is the spin for $\mathrm{Cr}^{3+}$ ion.
Using the expression for the potential energy Eq.~\eqref{eq:U}, one can derive the corresponding Lagrangian and solving the Euler-Lagrange equations describe the spin dynamics triggered by the electromagnetic field. In particular, for  $m_{x}$ one finds
\begin{equation}
\label{eq:mx_manuscript}
     \ddot{m}_{x} + \omega_{\mathrm{M}}^{2} m_{x} = \pm \omega_{\mathrm{A}} \omega_{\mathrm{ME}\perp} E^{\mathrm{THz}}_{x} + \gamma \omega_{\mathrm{A}} H^{\mathrm{THz}}_{x},
\end{equation}
where $\omega_{\mathrm{M}}$ is the frequency of the antiferromagnetic resonance, $\omega_{\mathrm{A}}$ is the coefficient related to the uniaxial magnetic anisotropy, $\omega_{\mathrm{ME}\perp}$ is the parameter associated with the magnetoelectric coefficient $\alpha_{\perp}$, and $\pm$ operator distinguishes the antiferromagnetic domains with mutually opposite antiferromagnetic N{\'e}el vectors $\mathbf{L}_{\uparrow}$ and $\mathbf{L}_{\downarrow}$.
In Eq.~\eqref{eq:mx_manuscript}, the first right-side term ($E^{\mathrm{THz}}_{x}$-term) corresponds to the magneto-electric torque, while the second term ($H^{\mathrm{THz}}_{x}$-term) results in the conventional Zeeman torque.
Correspondingly,  these geometries we will be called as
magnetoelectric ($\mathbf{E}^{\mathrm{THz}} \parallel x$) and Zeeman ($\mathbf{H}^{\mathrm{THz}} \parallel x$) geometries, respectively. If one assumes that the experimentally observed oscillations are proportional to $m_x$, the model predicts that (a) the oscillations can be excited in the both magnetoelectric and Zeeman geometries; (b) the amplitude of the oscillations scales linearly with either of the fields  ($E^{\mathrm{THz}}_{x}$ or $H^{\mathrm{THz}}_{x}$); (c) the phase of the oscillations shifts by $\pi$ upon rotation of the fields by $\pi$ $E^{\mathrm{THz}}_{x} \leftrightarrow - E^{\mathrm{THz}}_{x}$ and $H^{\mathrm{THz}}_{x} \leftrightarrow - H^{\mathrm{THz}}_{x}$; (d) the oscillations excited due to the magneto-electric torque do change sign upon reversal of the antiferromagnetic N{\'e}el vector $\mathbf{L}_{\uparrow}  \leftrightarrow \mathbf{L}_{\downarrow}$, while the oscillations excited via the Zeeman torque do not. The first three peculiarities do agree with those observed experimentally in the both magnetoelectric and Zeeman geometries. It is thus interesting to study how in our experiment the THz induced spin oscillations depend on the antiferromagnetic N{\'e}el vector. 

\begin{figure*}[ht]
\centering
\includegraphics[width=1\columnwidth]{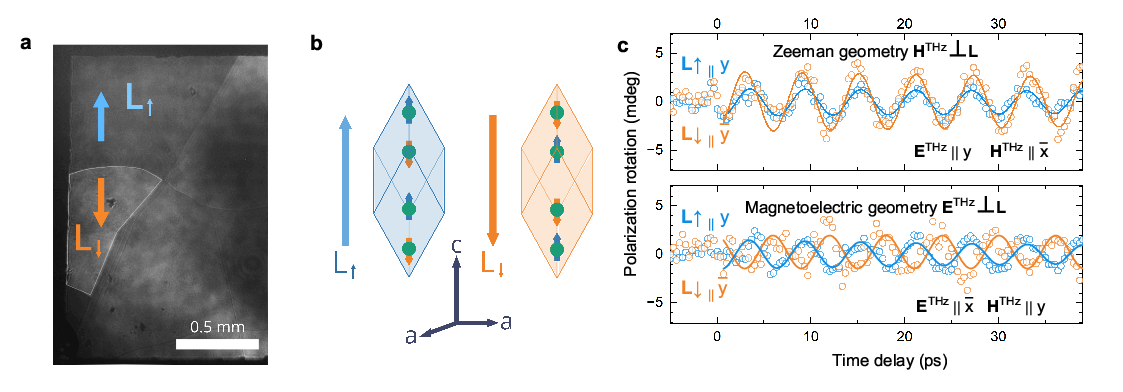}
\caption{\label{fig:2}
\textbf{a}~Antiferromagnetic domain pattern of the \CrO{} sample visualized be the technique of the second harmonic generation as described in Methods.
The domains with mutually opposite antiferromagnetic N{\'e}el vectors $\mathbf{L}_{\uparrow}$ (blue) and $\mathbf{L}_{\downarrow}$ (orange) are marked correspondingly. 
\textbf{b}~The rhombohedral unit cell of \CrO{} (only magnetic Cr$^{3+}$ ions are shown) with antiferromagnetic domains having opposite arrangement of spins and opposite N{\'e}el vectors $\leftarrow\,\rightarrow\;\;\leftarrow\,\rightarrow$ ($\mathbf{L}_{\uparrow}$) and $\rightarrow\,\leftarrow\;\;\rightarrow\,\leftarrow$ ($\mathbf{L}_{\downarrow}$), respectively.~\cite{brockhouse1953antiferromagnetic,corliss1965magnetic}.
\textbf{c}~Transient probe polarization rotation induced in the two domains with mutually opposite directed antiferromagnetic N{\'e}el vectors $\mathbf{L}_{\uparrow}$ (blue) and $\mathbf{L}_{\downarrow}$ (orange) for the Zeeman $\mathbf{H}^{\mathrm{THz}} \perp \mathbf{L}$ ($\mathbf{E}^{\mathrm{THz}} \parallel \mathbf{L}$) and the magnetoelectric $\mathbf{E}^{\mathrm{THz}} \perp \mathbf{L}$ ($\mathbf{H}^{\mathrm{THz}} \parallel  \mathbf{L}$) experimental geometry, respectively. The measurements are performed at $T = 77$ K.  
The circles and solid lines are experimental data points and guides for the eye, respectively.
}
\end{figure*}

Here we benefited from the possibility to visualize antiferromagnetic domains in \CrO{} (see Fig.~\ref{fig:2}a) and performed the experiments in the areas with two opposite orientations of the antiferromagnetic N{\'e}el vector $\mathbf{L}_{\uparrow}$ and $\mathbf{L}_{\downarrow}$. Fig.~~\ref{fig:2}b shows how the spins of four magnetic sublattices are aligned in these two types of domains. THz induced spin dynamics is shown in Fig.~\ref{fig:2}c and demonstrates a clear difference between the cases when the THz magnetic field $\mathbf{H}^{\mathrm{THz}}$ is parallel and perpendicular to the N{\'e}el vector $\mathbf{L}$ of the antiferromagnet, respectively.
When the THz magnetic field and the antiferromagnetic N{\'e}el vector are mutually perpendicular $\mathbf{H}^{\mathrm{THz}} \perp \mathbf{L}$ (Zeeman torque), a reversal of $\mathbf{L}$ does not affect the detected spin dynamics.  
On the contrary, when the THz magnetic field and the antiferromagnetic N{\'e}el vector are mutually parallel  $\mathbf{H}^{\mathrm{THz}} \parallel \mathbf{L}$ which is equivalent to $\mathbf{E}^{\mathrm{THz}} \perp \mathbf{L}$ , a reversal of $\mathbf{L}$ is accompanied a shift of the phase of the oscillations by $\pi$.
These experimental results are in perfect agreement with the dynamics predicted by Eq.~\eqref{eq:mx_manuscript}. All these experimental findings strongly suggest that next to the THz magnetic field   $\mathbf{H}^{\mathrm{THz}} $, spin dynamics in the antiferromagnet can be excited by the THz electric field   $\mathbf{E}^{\mathrm{THz}}$. In the particular case of fields of freely propagating electromagnetic plane wave, spin dynamics excited via these two mechanisms have comparable amplitudes. Interestingly, the amplitudes of the oscillations excited via these two mechanisms also have nearly similar temperature dependencies (Fig.\ref{fig:1}e). 

Finally, we note that $l_z$ component of the N{\'e}el vector may show up in our experiment similarly to $m_x$ component of the net magnetization. Both components can, in principle, contribute to the polarization rotation in our experiment. While $m_x$ can contribute to the magneto-optical Faraday effect if the probe beam propagates at an angle with the respect to the $z$-axis as a result of optical alignment or a miscut of the crystal,  $l_z$ can contribute to the polarization rotation due to magnetic linear dichroism in the $xy$-plane as discussed in details in Supplementary section.

\section*{CONCLUSIONS}
In conclusion, we demonstrated THz magneto-electric effect in \CrO{} showing that even if the electric field is applied at THz rates, it can still be efficiently transformed into effective THz magnetic field and thus launch spin resonance in the collinear antiferromagnet.
In the particular case of freely propagating THz plane wave,  the torques on the antiferromagnetic N{\'e}el vector induced by the magnetic and the electric fields are shown to have comparable effects on spin dynamics and similar temperature dependencies. Both effects are at maximum when the corresponding field (electric or magnetic) is perpendicular to the antiferromagnetic N{\'e}el vector. 
Our work is practically a game-changer in the coupling of THz fields to antiferromagnetic spins as it shows that even in collinear antiferromagnets spins can be controlled by picosecond pulses of electric field. Such pulses can be applied on chip with the help of electrodes even in nanospintronic devices thus benefiting from a much better spatial resolution than in the case of control with the help of magnetic fields. Also in magnonics, formerly well understood coupling of freely electromagnetic wave to an antiferromagnet via an antenna or metamaterial must be reconsidered. Not only evanescent magnetic, but also evanescent electric field will excite spins and generate spin waves suggesting that THz magnonics is much richer field than it has been believed until now. 
Hence, our discovery practically opens up new chapters in ultrafast magnetism, antiferromagnetic spintronics and THz magnonics.

\newpage

\section*{EXPERIMENTAL PROCEDURES / METHODS}

\subsection*{Cr$_{2}$O$_{3}$ material properties and sample information}
Chromium oxide \CrO{} is a prototypical magnetoelectric antiferromanget with the corundum trigonal crystal structure (s.g. $R\overline{3}c$).
We use the hexagonal setting of trigonal crystal system.
Below the N{\'e}el temperature $T_{N} = 307$\,K~\cite{volger1952anomalous,mcguire1956antiferromagnetism}, the four $\mathrm{Cr}^{3+}$ spins in the unit cell antiferromagnetically ordered alternating in an up and down sequence along the hexagonal $c$ axis (magnetic space group $R\overline{3}'c'$)~\cite{brockhouse1953antiferromagnetic,corliss1965magnetic}.
There are two types of opposite antiferromagnetic domains $\leftarrow\,\rightarrow\;\;\leftarrow\,\rightarrow$ ($\mathbf{L}_{\uparrow}$) and $\rightarrow\,\leftarrow\;\;\rightarrow\,\leftarrow$ ($\mathbf{L}_{\downarrow}$) in \CrO~\cite{corliss1965magnetic,bousquet2024sign}.
The spin structure of \CrO{} (magnetic point group $\underline{\overline{3}m}$) breaks inversion and time-reversal symmetries, while their combination is preserved which allows the linear magnetoelectric effect~\cite{dzyaloshinskii1960magneto}.

In the experiments we used a $10\overline{1}0$ single-crystal slab of \CrO{} with thickness 100\,$\mu$m of the optical quality of the surfaces which was purchased from MaTecK GmbH.
The magnetic easy axis is along the $c$ axis lying in the sample plane.

\subsection*{Experimental setup}
\begin{figure}[ht]
\centering
\includegraphics[width=1\columnwidth]{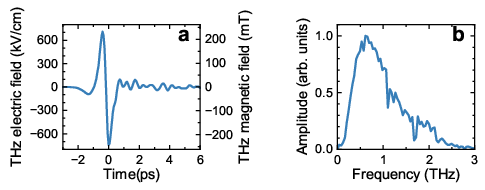}
\caption{\label{fig:S1}
\textbf{a}~Terahertz pulse waveform and \textbf{b}~corresponding normalized Fourier spectrum.
}
\end{figure}

To study the effect of THz electric field on spin dynamics, we performed THz pump-infrared probe experiments on \CrO{} sample.
The sample is placed in an open-cycle cryostat, which is cooled with liquid nitrogen. It allowed us to control the sample temperature from 77 to 300\,K.
The laser system operated at 1\,kHz repetition rate and provided pulse with duration of 100\,fs at the central wavelength of 800\,nm.
The linearly polarized THz pump pulses were generated by tilted-front optical rectification of laser pulses in $\mathrm{LiNbO}_{3}$ prism.
The peak electric field strength of the THz pulses by electro-optical sampling in GaP crystal of 50\,$\mu$m thickness was estimated as 760\,kV/cm and the corresponding magnetic field was 250\,mT.
The spectrum of the THz pump pulse with the maximum at 0.6\,THz is shown in Fig.\ref{fig:S1}b.
The orientation and the strength of the THz electric field  were controlled by two wire-grid polarizes. 
The THz pump and IR probe pulses were focused and spatially overlapped on the sample surface under close to normal incidence in the area of about 300 and 60\,$\mu$m, correspondingly.
We probe the THz induced spin dynamics magneto-optically by measuring the polarization rotation of the initially linearly polarized probe beam transmitted through the sample.
The experiments were performed in a dry ambient atmosphere to remove water and thus avoid THz absorption at the spectral lines of water.
To distinguish two types of antiferromagnetic domains, the second harmonic microscopy with elliptically polarized probe pulses was employed as described in Ref.~\cite{fiebig1995domain}.

\subsection*{Theoretical model and simulations}

To describe the observed spin dynamics we developed a model of magnetoelectric excitation of antiferromagnetic resonance in the antiferromagnet \CrO.
In the two-sublattice approximation the four magnetizations of alternating magnetic sublattices in \CrO{} are replaced by two opposite normalized magnetizations $\mathbf{m}_{\mathrm{A}} = (\mathbf{m}_{1} + \mathbf{m}_{3}) / 2$ and $\mathbf{m}_{\mathrm{B}} = (\mathbf{m}_{2} + \mathbf{m}_{4}) / 2$~\cite{turov2001symmetry}.
We define the net magnetization vector $\mathbf{m} = (\mathbf{m}_{\mathrm{A}} + \mathbf{m}_{\mathrm{B}}) / 2$ and antiferromagnetic N{\'e}el vector $\mathbf{l} = (\mathbf{m}_{\mathrm{A}} - \mathbf{m}_{\mathrm{B}}) / 2$.
The antiferromagnetic N{\'e}el vector $\mathbf{l}$ in our Cartesian coordinate system is directed along the $y$ axis.
In the polar coordinate system with polar $\vartheta$ and azimuthal $\varphi$ angles, the sublattice magnetization vectors are $\mathbf{m_{\mathrm{A}(\mathrm{B})}} = (\sin{\vartheta_{\mathrm{A}(\mathrm{B})}}\cos{\varphi_{\mathrm{A}(\mathrm{B})}}, \sin{\vartheta_{\mathrm{A}(\mathrm{B})}}\sin{\varphi_{\mathrm{A}(\mathrm{B})}}, \cos{\vartheta_{\mathrm{A}(\mathrm{B})}})$.
Then, $\mathbf{m}_{\mathrm{A}}$ and $\mathbf{m}_{\mathrm{B}}$ are parameterized by
\begin{equation}
\label{eq:canted_angles}
\begin{gathered}
    \vartheta_{\mathrm{A}} = \vartheta - \epsilon, \quad \vartheta_{\mathrm{B}} = \pi - \vartheta - \epsilon,\\
    \varphi_{\mathrm{A}} = \varphi + \beta, \quad \varphi_{\mathrm{B}} = \pi + \varphi - \beta,
\end{gathered}
\end{equation}
where small canting angles $\epsilon \ll 1$ and $\beta \ll 1$ are introduced. 
Expand the net magnetization $\mathbf{m}$ and antiferromagnetic N{\'e}el $\mathbf{l}$ vectors in series with regards to small canting angles $\epsilon$ and $\beta$, we obtain
\begin{equation}
\label{eq:vectors}
\begin{gathered}
    m_{x} \approx  - \beta \sin{\vartheta} \sin{\varphi} - \epsilon \cos{\vartheta} \cos{\varphi},\\
    m_{y} \approx \beta \sin{\vartheta} \cos{\varphi} - \epsilon \cos{\vartheta} \sin{\varphi},\\
    m_{z} \approx \epsilon \sin{\vartheta},\\
    l_{x} \approx \sin{\vartheta} \cos{\varphi},\\
    l_{y} \approx \sin{\vartheta} \sin{\varphi},\\
    l_{z} \approx \cos{\vartheta}.
\end{gathered}
\end{equation}

The ground state of $\mathbf{m}$ and $\mathbf{l}$ vectors in \CrO{} can be defined by two angles $\vartheta_{0} = \pi/2$ and $\varphi_{0} = \pm \pi/2$, where $\pm$ operator denotes two different antiferromagnetic domains. 
Near the ground state the angles can be expressed as $\vartheta = \vartheta_{0} + \vartheta_{1}$ and $\varphi = \varphi_{0} + \varphi_{1}$, where $\vartheta_{1} \ll 1$ and $\varphi_{1} \ll 1$.
Then, taking into account the following relations 
\begin{equation}
\begin{gathered}
    \sin{\vartheta} \approx 1 - \frac{\vartheta_{1}^2}{2}, \quad \cos{\vartheta} \approx - \vartheta_{1}, \\
    \sin{\varphi} \approx  \pm \left( 1 - \frac{\varphi_{1}^2}{2} \right), \quad \cos{\varphi} \approx \mp \varphi_{1},
\end{gathered}
\end{equation}
we can represent vector components $\mathbf{m}$ and $\mathbf{l}$ [Eqs.~\eqref{eq:vectors}] in the form
\begin{equation}
\label{eq:vectors_small}
\begin{gathered}
    m_{x} \approx  \mp \beta,\\
    m_{y} \approx \pm (\epsilon \vartheta_{1} - \beta \varphi_{1}),\\
    m_{z} \approx \epsilon,\\
    l_{x} \approx \mp \varphi_{1},\\
    l_{y} \approx \pm \left( 1 - \frac{\varphi_{1}^2}{2} - \frac{\vartheta_{1}^{2}}{2}\right), \\
    l_{z} \approx - \vartheta_{1}.
\end{gathered}
\end{equation}

The kinetic energy of the spin system of a double-sublattice antiferromagnet can be determined through the Berry phase gauge $\gamma_\mathrm{Berry} = (1 - \cos{\vartheta_{\mathrm{A}}})\dot{\varphi}_{\mathrm{A}} + (1 - \cos{\vartheta_{\mathrm{B}}})\dot{\varphi}_{\mathrm{B}}$~\cite{fradkin2013field,zvezdin2024giant} in the first order in $\epsilon$ and $\beta$ for \CrO{} as
\begin{equation}
\label{eq:T}
    T = \frac{M_{0}}{\gamma} \, (\epsilon \dot{\varphi}_{1} + \beta \dot{\vartheta}_{1}),
\end{equation}
where $M_{0}$ is the sublattice magnetization, $\gamma$ is the gyromagnetic ratio.

The exchange energy of the spin system of a double-sublattice antiferromagnet \CrO{} in the second order in $\epsilon$ and $\beta$ up to a constant term can be written as
\begin{equation}
\label{eq:J}
    U_{\mathrm{Ex}} = \frac{M_{0}^{2}}{\chi_{\perp}} \, \mathbf{m}_{\mathrm{A}} \cdot \mathbf{m}_{\mathrm{B}} \approx \frac{2M_{0}^{2}}{\chi_{\perp}} (\beta^{2} + \epsilon^{2}),
\end{equation}

where $\chi_{\perp}$ is the perpendicular magnetic susceptibility.

The uniaxial magnetic anisotropy of \CrO{} using Eq.~\eqref{eq:vectors_small} in the first order in $\varphi_{1}$ and $\vartheta_{1}$ up to a constant term has the following form
\begin{equation}
\label{eq:K}
    U_{\mathrm{A}} = - K l_{y}^{2} \approx K (\varphi_{1}^{2} + \vartheta_{1}^{2}),
\end{equation}
where $K$ is the uniaxial anisotropy constant. 

The magnetoelectric interaction with the electric field $\mathbf{E}$ applied in the $xy$ plane in \CrO{}, taking into account Eqs.~\eqref{eq:vectors_small}, has the following form~\cite{belov1993magnetoelectric}
\begin{equation}
\label{eq:UME}
    U_{\mathrm{ME}} = - \lambda_{\perp} \kappa_{\perp} M_{0} m_{x} l_{y} E_{x} - \lambda_{\parallel} \kappa_{\parallel} M_{0} m_{y} l_{y} E_{y} = [ \lambda_{\perp} \kappa_{\perp} M_{0} \beta E_{x} \,  - \lambda_{\parallel} \kappa_{\parallel} M_{0} (\epsilon \vartheta_{1} - \beta \varphi_{1}) E_{y} ],
\end{equation}
where $\lambda_{\perp(\parallel)}$ are parameters which are related to the magnetoelectric coefficients $\alpha_{\perp(\parallel)}$~\cite{astrov1961magnetoelectric}, $\kappa_{\perp(\parallel)} = \cfrac{\varepsilon_{\perp(\parallel)} - 1}{4 \pi}$ are the electric susceptibilities, and $\varepsilon_{\perp(\parallel)}$ are dielectric permittivities perpendicular and parallel to the optical axis, respectively.

The Zeeman interaction of the spin system with the magnetic field $\mathbf{H}$ applied in the $xy$ plane in \CrO{} can be represented as 
\begin{equation}
\label{eq:Z}
        U_{\mathrm{Z}} = - M_{0} \, \mathbf{m} \cdot \mathbf{H} = - M_{0} \left( m_{x} H_{x} + m_{y} H_{y} \right) =  \pm M_{0} \left[ \beta H_{x} + ( \beta \varphi_{1} - \epsilon \vartheta_{1} ) H_{y} \right].
\end{equation}

To describe the spin dynamics driven by the THz pulse we employ a Lagrangian
\begin{equation}
\begin{gathered}
\label{eq:L}
    \mathcal{L} = T - U_{\mathrm{Ex}} - U_{\mathrm{A}} - U_{\mathrm{ME}} - U_{\mathrm{Z}}  = \frac{M_{0}}{\gamma} \, (\epsilon \dot{\varphi}_{1} + \beta \dot{\vartheta}_{1}) - \frac{2M_{0}^{2}}{\chi_{\perp}} \, (\beta^{2} + \epsilon^{2}) - K (\varphi_{1}^{2} + \vartheta_{1}^{2}) \\ - \left[ \lambda_{\perp} \kappa_{\perp} M_{0} \beta E_{x} - \lambda_{\parallel} \kappa_{\parallel} M_{0} (\epsilon \vartheta_{1} - \beta \varphi_{1}) E_{y} \right] \mp M_{0} \left[ \beta H_{x} + ( \beta \varphi_{1} - \epsilon \vartheta_{1} ) H_{y} \right].
\end{gathered}
\end{equation}
Note that all terms in Eq.~\eqref{eq:L} are considered for a single molecule units.  
Then we insert the Lagrangian~\eqref{eq:L} into the Euler-Lagrange equations
\begin{equation}
    \frac{d}{dt} \frac{\partial \mathcal L}{\partial \dot q_{i}} - \frac{\partial \mathcal L}{\partial q_{i}} = 0,
\end{equation}
where $q_{i}$ for $i=1$--$4$ are order parameters $\epsilon$, $\varphi_{1}$, $\beta$, and $\vartheta_{1}$, respectively.
As a result, we obtain a system of four coupled differential equations describing the spin dynamics of the magnetoelectric \CrO{} induced by the electric $\mathbf{E}$ and magnetic $\mathbf{H}$ fields of THz pulse
\begin{equation}
\label{eq:diff_eq}
\begin{aligned}
    &\dot{\epsilon} + \omega_{\mathrm{A}} \varphi_{1} + \omega_{\mathrm{ME}\parallel} \beta E_{y} \pm \gamma \beta H_{y} = 0, \\
    &\dot{\varphi}_{1} - \omega_{\mathrm{Ex}} \epsilon + \omega_{\mathrm{ME}\parallel} \vartheta_{1} E_{y} \pm \gamma \vartheta_{1} H_{y} = 0, \\
    &\dot{\beta} + \omega_{\mathrm{A}} \vartheta_{1} - \omega_{\mathrm{ME}\parallel} \epsilon E_{y} \mp \gamma \epsilon H_{y} = 0, \\
    &\dot{\vartheta}_{1} - \omega_{\mathrm{Ex}} \beta  - \omega_{\mathrm{ME}\parallel} \varphi_{1} E_{y} \mp \gamma \varphi_{1} H_{y} = \omega_{\mathrm{ME}\perp} E_{x} \pm \gamma H_{x},
\end{aligned}
\end{equation}
where introduced constant are $\omega_{\mathrm{A}} = \gamma\;H_{\mathrm{A}} = 2 \gamma K / M_{0}$, $\omega_{\mathrm{Ex}} = \gamma\;H_{\mathrm{Ex}} = 4 \gamma M_{0} / \chi_{\perp}$, $\omega_{\mathrm{ME}\perp} = \gamma \lambda_{\perp} \kappa_{\perp}$, and $\omega_{\mathrm{ME}\parallel} = \gamma \lambda_{\parallel} \kappa_{\parallel}$.

The Floquet system of differential equations~\eqref{eq:diff_eq} describes the dynamics of a doubly degenerated magnon with the frequency $\omega_{\mathrm{M}} = \sqrt{\omega_{\mathrm{Ex}}\omega_{\mathrm{A}}}$ which has been excited by the magnetoelectric ($E_{x}$-term) and Zeeman ($H_{x}$-term) torques.  
Besides, there are internal parametric magnetoelectric ($E_{y}$-term) and Zeeman ($H_{y}$-term)  torques which can show up significantly under conditions of parametric instability.

For the numerical simulation of spin dynamics we solve Eqs.~\eqref{eq:diff_eq} assuming that the THz electric field is defined as 
\begin{equation}
    \label{eq:E_THz}
    E^{\mathrm{THz}}(t) = -E_{0} \exp{\left({-\frac{t^{2}}{\tau_{\mathrm{THz}}^{2}}} \right)} \sin{\omega_{\mathrm{THz}}t},
\end{equation}
where $E_{0}$ is the peak electric field strength, $\tau_{\mathrm{THz}}$ and $\omega_{\mathrm{THz}}$ determine the THz pulse duration and frequency, respectively.
Note that, in the particular case of electromagnetic plane wave the  magnetic field $H^{\mathrm{THz}}(t)$ is related to the electric field $E^{\mathrm{THz}}(t)$   $E^{\mathrm{THz}}(t) = c H^{\mathrm{THz}}(t)$, where $c$ is the speed of light in vacuum.
For the THz pulse propagating along the $z$ axis the electric and magnetic field strength vectors are defined as $\mathbf{E}^{\mathrm{THz}} = E^{\mathrm{THz}}(t)(\cos{\alpha}, \sin{\alpha}, 0)$ and $\mathbf{H}^{\mathrm{THz}} = H^{\mathrm{THz}}(t)(-\sin{\alpha}, \cos{\alpha}, 0)$, where $\alpha$ is the polarization angle with respect to the $x$ axis.

\begin{figure}[ht]
\centering
\includegraphics[width=1\columnwidth]{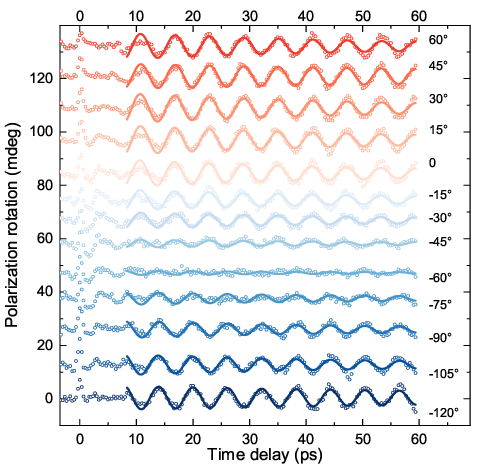}
\caption{\label{fig:S3}
Transient probe polarization rotation measured at different angles of the THz electric field  with respect to the $x$ axis in a single antiferromagnetic domain at $T = 77$\,K.
}
\end{figure}

To estimate the magnon damping time, we employ the linewidth of the magnon infrared absorption $\Delta\omega \approx 1.4$\,GHz from Ref.~\cite{mukhin1997bwo}.
Therefore, the magnon damping time is about $\tau \approx 725$\,ps that significantly exceeds the delay time range used in our experiments (about 100\,ps) at which the magnon oscillations are practically not dumped (see Fig.~\ref{fig:S3}).
Hence, we do not take into account the magnon damping in our simulations of spin dynamics.

The results of numerical simulation of spin dynamics from Eqs.~\eqref{eq:diff_eq} with the THz pump pulse from Eq.~\eqref{eq:E_THz} showed that the observed dynamics is likely due to  $m_{x}$ since the other components $m_{y}$ and $m_{z}$ are many orders of magnitude smaller. 
Besides, it turns out that the internal parametric magnetoelectic ($E_{y}$-term) and Zeeman ($H_{y}$-term) torques from Eqs.~\eqref{eq:diff_eq} can be neglected in a fair description of the obtained results for the electric and magnetic field strengths used in the experiment. 
Then the system of Eqs.~\eqref{eq:diff_eq} reduced to one second order differential equation for $\beta$ with two torques 
\begin{equation}
\label{eq:beta}
     \ddot{\beta} + \omega_{\mathrm{M}}^{2} \beta = - \omega_{\mathrm{A}} \omega_{\mathrm{ME}\perp} E^{\mathrm{THz}}_{x} \mp \gamma \omega_{\mathrm{A}} H^{\mathrm{THz}}_{x}.
\end{equation}
Therefore, it follows from Eqs.~\eqref{eq:vectors_small} and~\eqref{eq:beta} that the magnetic dynamics of $m_{x}$ obeys to the differential equation
\begin{equation}
\label{eq:mx}
     \ddot{m}_{x} + \omega_{\mathrm{M}}^{2} m_{x} = \pm \omega_{\mathrm{A}} \omega_{\mathrm{ME}\perp} E^{\mathrm{THz}}_{x} + \gamma \omega_{\mathrm{A}} H^{\mathrm{THz}}_{x}.
\end{equation}

For the projection of the antiferromagnetic N{\'e}el vector on the $z$ axis $l_{z} \approx - \vartheta_{1}$ [Eq.~\eqref{eq:vectors_small}] the equation of motion derived from the system of Eqs.~\eqref{eq:diff_eq} and neglecting the terms with $E_{y}$ and $H_{y}$ has the following form
\begin{equation}
\label{eq:lz}
     \ddot{l_{z}} + \omega_{\mathrm{M}}^{2} l_{z} = - \omega_{\mathrm{ME}\perp} \dot{E}^{\mathrm{THz}}_{x} \mp \gamma \dot{H}^{\mathrm{THz}}_{x}.
\end{equation}
According to this, the magnetoelectric torque of the antiferromagnetic N{\'e}el vector $\mathbf{l}$ is proportional to the time-derivative of the THz electric field $\propto \dot{E}^{\mathrm{THz}}_{x}$.
It is worth noting that from the $\pm$ sign in the right side of Eqs.~\eqref{eq:mx} and~\eqref{eq:lz} it follows that moving to the opposite antiferromagnetic domain differently shifts phase by $\pi$ for $m_{x}$ and $l_{z}$.
So, in the magnetoelectric geometry ($E_{x}$ term) the phase of $m_{x}$ is shifted, whereas in the Zeeman geometry ($H_{x}$ term) it is to be expected for $l_{z}$.
However, the combination of $l_{y} l_{z}$ has a linear dependence on the THz field strength and behaves like $m_{x}$ upon switching between two types of antiferromagnetic domains.

Now we can attribute the magnetoelectric coefficient $\alpha_{\perp}$ to the parameter $\lambda_{\perp}$ from Eq.~\eqref{eq:UME}.
By definition, $\alpha_{\perp}$ couples the induced magnetization to the applied electric field $M_{x} = \alpha_{\perp} E_{x}$.
The magnetization $M_{x} = M_{0} \, m_{x}$ can be obtained by solving Eq.~\eqref{eq:mx}, which in the static case has the form
\begin{equation}
\label{eq:Mx_static}
     M_{x} = \pm \lambda_{\perp} \kappa_{\perp} \chi_{\perp} E_{x} + \chi_{\perp} H_{x}.
\end{equation}
Therefore, the magnetoelectric coefficient $\alpha_{\perp}$ in the static case is expressed as
\begin{equation}
\label{eq:alpha}
     \alpha_{\perp} = \pm \lambda_{\perp} \kappa_{\perp} \chi_{\perp}.
\end{equation}
Note that the magnetoelectric coefficient $\alpha_{\perp}$ experiences a sign upon switching between the two antiferromagnetic domains, which is in a fair agreement with Ref.~\cite{bousquet2024sign}. 
Using the static values of $\kappa_{\perp} = 0.7$, $\chi_{\perp} = 1.2 \  10^{-4}$ and $\alpha_{\perp} = - 9 \  10^{-5}$ from Refs.~\cite{mukhin1997bwo,foner1963high,astrov1961magnetoelectric}, we can estimate the static parameters as $\lambda_{\perp}\kappa_{\perp} \approx -0.8$ and $\lambda_{\perp} \approx -1.1$.
Expanding the definition of static magnetoelectric coefficient $\alpha_{\perp}$~\eqref{eq:alpha} from static to THz case and considering the proximity of the magnetoelectric and Zeeman torques from Eq.~\eqref{eq:mx} $|\omega_{\mathrm{A}} \omega_{\mathrm{ME}\perp} E^{\mathrm{THz}}_{x}| \approx |\gamma \omega_{\mathrm{A}} H^{\mathrm{THz}}_{x}|$, we can estimate the THz parameters $\lambda^{\mathrm{THz}}_{\perp}\kappa^{\mathrm{THz}}_{\perp} = H^{\mathrm{THz}}_{x} / E^{\mathrm{THz}}_{x} \approx -1$, $\lambda^{\mathrm{THz}}_{\perp} \approx -1.4$, and $\alpha^{\mathrm{THz}}_{\perp} \approx \chi^{\mathrm{THz}}_{\perp} \approx -1.2 \  10^{-4}$. 

Next, numerically solving the system of differential equations~\eqref{eq:diff_eq} with the exchange frequency $\omega_{\mathrm{Ex}} = \gamma \, H_\mathrm{Ex} \approx 8.6~10^{13}$\,rad/s and the anisotropy frequency $\omega_{\mathrm{A}} = \gamma \, H_\mathrm{A} \approx 1.2~10^{10}$\,rad/s from Ref.~\cite{foner1963high}, we can estimate the small angles $\beta$ and $\vartheta_{1}$ [Eq.~\eqref{eq:vectors_small}] induced by the THz pump pulse [Eq.~\eqref{eq:E_THz}], in our experiment.
The amplitude of the spin canting angle oscillations is about $\beta \approx 0.0075^{\circ}$, while the amplitude of the deflection of the antiferromagnetic N{\'e}el vector is about $\vartheta_{1} \approx 0.6^{\circ}$.

\subsection*{Detection of spin dynamics}
In the ground state the antiferromagnetic N{\'e}el vector $\mathbf{L}$ is along the $y$-axis.
Both magnetoelectric torque and the Zeeman torque tilt the spins in the $yz$-plane.
In the following time domain, the spins aim to return to the ground state via damped oscillation around the equilibrium orientation.
In this case, contrary to ferromagnet, oscillations of the antiferromagnetic spins towards the ground state should occur essentially in the $yz$-plane.
It means that during the oscillations $l_{z}$\(\gg\)$l_{x}$.
Moreover, even if in the ground state the net magnetization of the antiferromagnet is zero, coherent oscillations of the spins in the $yz$-plane must induce oscillating magnetization along the $x$-axis 
($\mathbf{m} \propto [\mathbf{l}\times\dot{\mathbf{l}}]$
i.e. $m_{x} \propto l_{z}\,l_{y}$).
It means that the oscillations can result in the detected polarization rotation of the probe beam via two mechanisms.
Due to a miscut in the crystal giving that the wave-vector of the probe beam and the induced magnetization are not exactly perpendicular ($\mathbf{k}\cdot\mathbf{m}\neq0$) the spin oscillations will result in polarization rotation due to magneto-optical Faraday effect.
Alternatively, the signal can originate from the magnetic linear dichroism i.e. optical anisotropy induced in the $xy$ plane by spin oscillations~\cite{nemec2018antiferromagnetic}.
If $\varepsilon_{ij}$ the dielectric permittivity tensor, the anisotropy can be expressed by the difference $\varepsilon_{xx} -\varepsilon_{yy} \neq 0$, where $\varepsilon_{xx} \propto \zeta_{xxyz} l_{y} l_{z}$~\cite{birss1964symmetry}.

\newpage

\subsection*{Lead contact}

Requests for further information and resources should be directed to and will be fulfilled by the lead contact, Vladislav Bilyk (vladislav.bilyk@ru.nl).

\subsection*{Materials availability}

This study did not generate new materials.

\subsection*{Data and code availability}

Any additional information required to reanalyze the data reported in this paper is available from the lead contact upon request.    

\section*{ACKNOWLEDGMENTS}

The work was supported by the European Research Council ERC Grant Agreement number 101054664 (SPARTACUS) and the Gravitation program of the Dutch Ministry of Education, Culture and Science (OCW) under the research program “Materials for the Quantum Age” (QuMat) registration number 024.005.006.
R.\,M.\,D. acknowledges support of RSF (Grant No. 24-12-00348).
A.\,K.\,Z. acknowledges support of RSF (Grant No. 22-12-00367).

\section*{AUTHOR CONTRIBUTIONS}

Conceptualization, A.\,V.\,K.; methodology, V.\,R.\,B., A.\,V.\,K.; investigation,  V.\,R.\,B., R.\,M.\,D. and A.\,K.\,Z.; writing-–original draft, V.\,R.\,B. and R.\,M.\,D.; writing-–review \& editing, A.\,V.\,K., A.\,I.\,K. A.\,K.\,Z.; funding acquisition, A.\,V.\,K., R.\,M.\,D. and A.\,K.\,Z.; resources, R.\,M.\,D., A.\,K.\,Z., A.\,I.\,K.; supervision, A.\,V.\,K.

\section*{DECLARATION OF INTERESTS}

The authors declare that this work has been published as a result of peer-to-peer scientific collaboration between researchers.
The provided affiliations represent the actual addresses of the authors in agreement with their digital identifier (ORCID) and cannot be considered as a formal collaboration between the aforementioned institutions.

\newpage

\section*{MAIN FIGURE TITLES AND LEGENDS}

\newpage

\newpage

\newpage

\newpage

\bigskip


\newpage

\begin{thebibliography}{47}
\providecommand{\natexlab}[1]{#1}
\providecommand{\url}[1]{\texttt{#1}}
\providecommand{\href}[2]{#2}
\providecommand{\path}[1]{#1}
\providecommand{\DOIprefix}{doi:}
\providecommand{\ArXivprefix}{arXiv:}
\providecommand{\URLprefix}{URL: }
\providecommand{\Pubmedprefix}{pmid:}
\providecommand{\doi}[1]{\href{http://dx.doi.org/#1}{\path{#1}}}
\providecommand{\Pubmed}[1]{\href{pmid:#1}{\path{#1}}}
\providecommand{\BIBand}{and}
\providecommand{\bibinfo}[2]{#2}
\ifx\xfnm\undefined \def\xfnm[#1]{\unskip,\space#1}\fi
\makeatletter\def\@biblabel#1{#1.}\makeatother
\bibitem[{Yang et~al.(2023)Yang, Li, Fiebig and Pal}]{yang2023terahertz}
\bibinfo{author}{Yang, C.-J.}, \bibinfo{author}{Li, J.}, \bibinfo{author}{Fiebig, M.}, and \bibinfo{author}{Pal, S.} (\bibinfo{year}{2023}). \bibinfo{title}{Terahertz control of many-body dynamics in quantum materials}.
\newblock \bibinfo{journal}{Nat. Rev. Mater.} \emph{\bibinfo{volume}{8}}, \bibinfo{pages}{518--532}. \URLprefix \url{https://doi.org/10.1038/s41578-023-00566-w}. \DOIprefix\doi{10.1038/s41578-023-00566-w}.
\bibitem[{Mostovoy(2024)}]{mostovoy2024multiferroics}
\bibinfo{author}{Mostovoy, M.} (\bibinfo{year}{2024}). \bibinfo{title}{Multiferroics: different routes to magnetoelectric coupling}.
\newblock \bibinfo{journal}{npj Spintronics} \emph{\bibinfo{volume}{2}}, \bibinfo{pages}{18}. \URLprefix \url{https://doi.org/10.1038/s44306-024-00021-8}. \DOIprefix\doi{10.1038/s44306-024-00021-8}.
\bibitem[{Hassanpour et~al.(2022)Hassanpour, Zemp, Tokunaga, Taguchi, Tokura, Lottermoser, Fiebig and Weber}]{hassanpour2022magnetoelectric}
\bibinfo{author}{Hassanpour, E.}, \bibinfo{author}{Zemp, Y.}, \bibinfo{author}{Tokunaga, Y.}, \bibinfo{author}{Taguchi, Y.}, \bibinfo{author}{Tokura, Y.}, \bibinfo{author}{Lottermoser, T.}, \bibinfo{author}{Fiebig, M.}, and \bibinfo{author}{Weber, M.~C.} (\bibinfo{year}{2022}). \bibinfo{title}{Magnetoelectric transfer of a domain pattern}.
\newblock \bibinfo{journal}{Science} \emph{\bibinfo{volume}{377}}, \bibinfo{pages}{1109--1112}. \URLprefix \url{https://doi.org/10.1126/science.abm3058}. \DOIprefix\doi{10.1126/science.abm3058}.
\bibitem[{Leo et~al.(2018)Leo, Carolus, White, Kenzelmann, Hudl, Toledano, Honda, Kimura, Ivanov, Weil et~al.}]{leo2018magnetoelectric}
\bibinfo{author}{Leo, N.}, \bibinfo{author}{Carolus, V.}, \bibinfo{author}{White, J.~S.}, \bibinfo{author}{Kenzelmann, M.}, \bibinfo{author}{Hudl, M.}, \bibinfo{author}{Toledano, P.}, \bibinfo{author}{Honda, T.}, \bibinfo{author}{Kimura, T.}, \bibinfo{author}{Ivanov, S.~A.}, \bibinfo{author}{Weil, M.} et~al. (\bibinfo{year}{2018}). \bibinfo{title}{Magnetoelectric inversion of domain patterns}.
\newblock \bibinfo{journal}{Nature} \emph{\bibinfo{volume}{560}}, \bibinfo{pages}{466--470}. \URLprefix \url{https://doi.org/10.1038/s41586-018-0432-4}. \DOIprefix\doi{10.1038/s41586-018-0432-4}.
\bibitem[{Matsukura et~al.(2015)Matsukura, Tokura and Ohno}]{matsukura2015control}
\bibinfo{author}{Matsukura, F.}, \bibinfo{author}{Tokura, Y.}, and \bibinfo{author}{Ohno, H.} (\bibinfo{year}{2015}). \bibinfo{title}{Control of magnetism by electric fields}.
\newblock \bibinfo{journal}{Nat. Nanotech.} \emph{\bibinfo{volume}{10}}, \bibinfo{pages}{209--220}. \URLprefix \url{https://doi.org/10.1038/nnano.2015.22}. \DOIprefix\doi{10.1038/nnano.2015.22}.
\bibitem[{Lottermoser et~al.(2004)Lottermoser, Lonkai, Amann, Hohlwein, Ihringer and Fiebig}]{lottermoser2004magnetic}
\bibinfo{author}{Lottermoser, T.}, \bibinfo{author}{Lonkai, T.}, \bibinfo{author}{Amann, U.}, \bibinfo{author}{Hohlwein, D.}, \bibinfo{author}{Ihringer, J.}, and \bibinfo{author}{Fiebig, M.} (\bibinfo{year}{2004}). \bibinfo{title}{Magnetic phase control by an electric field}.
\newblock \bibinfo{journal}{Nature} \emph{\bibinfo{volume}{430}}, \bibinfo{pages}{541--544}. \URLprefix \url{https://doi.org/10.1038/nature02728}. \DOIprefix\doi{10.1038/nature02728}.
\bibitem[{Wu et~al.(2023)Wu, Bao, Zhou, Wang, Sun, Wen, Wan and Zhang}]{wu2023fluctuation}
\bibinfo{author}{Wu, F.}, \bibinfo{author}{Bao, S.}, \bibinfo{author}{Zhou, J.}, \bibinfo{author}{Wang, Y.}, \bibinfo{author}{Sun, J.}, \bibinfo{author}{Wen, J.}, \bibinfo{author}{Wan, Y.}, and \bibinfo{author}{Zhang, Q.} (\bibinfo{year}{2023}). \bibinfo{title}{Fluctuation-enhanced phonon magnetic moments in a polar antiferromagnet}.
\newblock \bibinfo{journal}{Nat. Phys.} \emph{\bibinfo{volume}{19}}, \bibinfo{pages}{1868--1875}. \URLprefix \url{https://doi.org/10.1038/s41567-023-02210-4}. \DOIprefix\doi{10.1038/s41567-023-02210-4}.
\bibitem[{Verseils et~al.(2023)Verseils, Hemme, Bounoua, Cervasio, Brubach, Houver, Gallais, Sacuto, Colson, Iijima et~al.}]{verseils2023stabilizing}
\bibinfo{author}{Verseils, M.}, \bibinfo{author}{Hemme, P.}, \bibinfo{author}{Bounoua, D.}, \bibinfo{author}{Cervasio, R.}, \bibinfo{author}{Brubach, J.}, \bibinfo{author}{Houver, S.}, \bibinfo{author}{Gallais, Y.}, \bibinfo{author}{Sacuto, A.}, \bibinfo{author}{Colson, D.}, \bibinfo{author}{Iijima, T.} et~al. (\bibinfo{year}{2023}). \bibinfo{title}{Stabilizing electromagnons in $\mathrm{CuO}$ under pressure}.
\newblock \bibinfo{journal}{npj Quantum Mater.} \emph{\bibinfo{volume}{8}}, \bibinfo{pages}{11}. \URLprefix \url{https://doi.org/10.1038/s41535-023-00542-1}. \DOIprefix\doi{10.1038/s41535-023-00542-1}.
\bibitem[{Masuda et~al.(2021)Masuda, Kaneko, Tokura and Takahashi}]{masuda2021electric}
\bibinfo{author}{Masuda, R.}, \bibinfo{author}{Kaneko, Y.}, \bibinfo{author}{Tokura, Y.}, and \bibinfo{author}{Takahashi, Y.} (\bibinfo{year}{2021}). \bibinfo{title}{Electric field control of natural optical activity in a multiferroic helimagnet}.
\newblock \bibinfo{journal}{Science} \emph{\bibinfo{volume}{372}}, \bibinfo{pages}{496--500}. \URLprefix \url{https://doi.org/10.1126/science.aaz4312}. \DOIprefix\doi{10.1126/science.aaz4312}.
\bibitem[{T{\'o}th et~al.(2016)T{\'o}th, Wehinger, Rolfs, Birol, Stuhr, Takatsu, Kimura, Kimura, R{\o}nnow and R{\"u}egg}]{toth2016electromagnon}
\bibinfo{author}{T{\'o}th, S.}, \bibinfo{author}{Wehinger, B.}, \bibinfo{author}{Rolfs, K.}, \bibinfo{author}{Birol, T.}, \bibinfo{author}{Stuhr, U.}, \bibinfo{author}{Takatsu, H.}, \bibinfo{author}{Kimura, K.}, \bibinfo{author}{Kimura, T.}, \bibinfo{author}{R{\o}nnow, H.~M.}, and \bibinfo{author}{R{\"u}egg, C.} (\bibinfo{year}{2016}). \bibinfo{title}{Electromagnon dispersion probed by inelastic {X}-ray scattering in $\mathrm{LiCrO}_{2}$}.
\newblock \bibinfo{journal}{Nat. Commun.} \emph{\bibinfo{volume}{7}}, \bibinfo{pages}{13547}. \URLprefix \url{https://doi.org/10.1038/ncomms13547}. \DOIprefix\doi{10.1038/ncomms13547}.
\bibitem[{Jones et~al.(2014)Jones, Gaw, Doig, Prabhakaran, H{\'e}troy~Wheeler, Boothroyd and Lloyd-Hughes}]{jones2014high}
\bibinfo{author}{Jones, S. P.~P.}, \bibinfo{author}{Gaw, S.~M.}, \bibinfo{author}{Doig, K.~I.}, \bibinfo{author}{Prabhakaran, D.}, \bibinfo{author}{H{\'e}troy~Wheeler, E.~M.}, \bibinfo{author}{Boothroyd, A.~T.}, and \bibinfo{author}{Lloyd-Hughes, J.} (\bibinfo{year}{2014}). \bibinfo{title}{High-temperature electromagnons in the magnetically induced multiferroic cupric oxide driven by intersublattice exchange}.
\newblock \bibinfo{journal}{Nat. Commun.} \emph{\bibinfo{volume}{5}}, \bibinfo{pages}{3787}. \URLprefix \url{https://doi.org/10.1038/ncomms4787}. \DOIprefix\doi{10.1038/ncomms4787}.
\bibitem[{Kubacka et~al.(2014)Kubacka, Johnson, Hoffmann, Vicario, De~Jong, Beaud, Gr{\"u}bel, Huang, Huber, Patthey et~al.}]{kubacka2014large}
\bibinfo{author}{Kubacka, T.}, \bibinfo{author}{Johnson, J.~A.}, \bibinfo{author}{Hoffmann, M.~C.}, \bibinfo{author}{Vicario, C.}, \bibinfo{author}{De~Jong, S.}, \bibinfo{author}{Beaud, P.}, \bibinfo{author}{Gr{\"u}bel, S.}, \bibinfo{author}{Huang, S.-W.}, \bibinfo{author}{Huber, L.}, \bibinfo{author}{Patthey, L.} et~al. (\bibinfo{year}{2014}). \bibinfo{title}{{L}arge-{A}mplitude {S}pin {D}ynamics {D}riven by a {THz} {P}ulse in {R}esonance with an {E}lectromagnon}.
\newblock \bibinfo{journal}{Science} \emph{\bibinfo{volume}{343}}, \bibinfo{pages}{1333--1336}. \URLprefix \url{https://doi.org/10.1126/science.1242862}. \DOIprefix\doi{10.1126/science.1242862}.
\bibitem[{Pimenov et~al.(2006)Pimenov, Mukhin, Ivanov, Travkin, Balbashov and Loidl}]{pimenov2006possible}
\bibinfo{author}{Pimenov, A.}, \bibinfo{author}{Mukhin, A.~A.}, \bibinfo{author}{Ivanov, V.~Y.}, \bibinfo{author}{Travkin, V.~D.}, \bibinfo{author}{Balbashov, A.~M.}, and \bibinfo{author}{Loidl, A.} (\bibinfo{year}{2006}). \bibinfo{title}{Possible evidence for electromagnons in multiferroic manganites}.
\newblock \bibinfo{journal}{Nature Phys.} \emph{\bibinfo{volume}{2}}, \bibinfo{pages}{97--100}. \URLprefix \url{https://doi.org/10.1038/nphys212}. \DOIprefix\doi{10.1038/nphys212}.
\bibitem[{Gao et~al.(2024)Gao, Peng, Cheng, Vi{\~n}as~Bostr{\"o}m, Kim, Jain, Vishnu, Raju, Sankar, Lee et~al.}]{gao2024giant}
\bibinfo{author}{Gao, F.~Y.}, \bibinfo{author}{Peng, X.}, \bibinfo{author}{Cheng, X.}, \bibinfo{author}{Vi{\~n}as~Bostr{\"o}m, E.}, \bibinfo{author}{Kim, D.~S.}, \bibinfo{author}{Jain, R.~K.}, \bibinfo{author}{Vishnu, D.}, \bibinfo{author}{Raju, K.}, \bibinfo{author}{Sankar, R.}, \bibinfo{author}{Lee, S.-F.} et~al. (\bibinfo{year}{2024}). \bibinfo{title}{{G}iant chiral magnetoelectric oscillations in a van der {W}aals multiferroic}.
\newblock \bibinfo{journal}{Nature}. \URLprefix \url{https://doi.org/10.1038/s41586-024-07678-5}. \DOIprefix\doi{10.1038/s41586-024-07678-5}.
\bibitem[{Han et~al.(2023)Han, Cheng, Liu, Ohno and Fukami}]{han2023coherent}
\bibinfo{author}{Han, J.}, \bibinfo{author}{Cheng, R.}, \bibinfo{author}{Liu, L.}, \bibinfo{author}{Ohno, H.}, and \bibinfo{author}{Fukami, S.} (\bibinfo{year}{2023}). \bibinfo{title}{Coherent antiferromagnetic spintronics}.
\newblock \bibinfo{journal}{Nat. Mater.} \emph{\bibinfo{volume}{22}}, \bibinfo{pages}{684--695}. \URLprefix \url{https://doi.org/10.1038/s41563-023-01492-6}. \DOIprefix\doi{10.1038/s41563-023-01492-6}.
\bibitem[{N{\v{e}}mec et~al.(2018)N{\v{e}}mec, Fiebig, Kampfrath and Kimel}]{nemec2018antiferromagnetic}
\bibinfo{author}{N{\v{e}}mec, P.}, \bibinfo{author}{Fiebig, M.}, \bibinfo{author}{Kampfrath, T.}, and \bibinfo{author}{Kimel, A.~V.} (\bibinfo{year}{2018}). \bibinfo{title}{Antiferromagnetic opto-spintronics}.
\newblock \bibinfo{journal}{Nat. Phys.} \emph{\bibinfo{volume}{14}}, \bibinfo{pages}{229--241}. \URLprefix \url{https://doi.org/10.1038/s41567-018-0051-x}. \DOIprefix\doi{10.1038/s41567-018-0051-x}.
\bibitem[{Baltz et~al.(2018)Baltz, Manchon, Tsoi, Moriyama, Ono and Tserkovnyak}]{baltz2018antiferromagnetic}
\bibinfo{author}{Baltz, V.}, \bibinfo{author}{Manchon, A.}, \bibinfo{author}{Tsoi, M.}, \bibinfo{author}{Moriyama, T.}, \bibinfo{author}{Ono, T.}, and \bibinfo{author}{Tserkovnyak, Y.} (\bibinfo{year}{2018}). \bibinfo{title}{Antiferromagnetic spintronics}.
\newblock \bibinfo{journal}{Rev. Mod. Phys.} \emph{\bibinfo{volume}{90}}, \bibinfo{pages}{015005}. \URLprefix \url{https://link.aps.org/doi/10.1103/RevModPhys.90.015005}. \DOIprefix\doi{10.1103/RevModPhys.90.015005}.
\bibitem[{Jungwirth et~al.(2016)Jungwirth, Marti, Wadley and Wunderlich}]{jungwirth2016antiferromagnetic}
\bibinfo{author}{Jungwirth, T.}, \bibinfo{author}{Marti, X.}, \bibinfo{author}{Wadley, P.}, and \bibinfo{author}{Wunderlich, J.} (\bibinfo{year}{2016}). \bibinfo{title}{Antiferromagnetic spintronics}.
\newblock \bibinfo{journal}{Nat. Nanotech.} \emph{\bibinfo{volume}{11}}, \bibinfo{pages}{231--241}. \URLprefix \url{https://doi.org/10.1038/nnano.2016.18}. \DOIprefix\doi{10.1038/nnano.2016.18}.
\bibitem[{Qiu et~al.(2023)Qiu, Tzschaschel, Ahn, Gao, Li, Zhang, Ghosh, Hu, Wang, Liu, B{\'e}rub{\'e}, Dinh, Gong, Lien, Ho, Singh, Watanabe, Taniguchi, Bell, Lu, Bansil, Lin, Chang, Zhou, Ma, Vishwanath, Ni and Xu}]{Qiu2023Axion}
\bibinfo{author}{Qiu, J.-X.}, \bibinfo{author}{Tzschaschel, C.}, \bibinfo{author}{Ahn, J.}, \bibinfo{author}{Gao, A.}, \bibinfo{author}{Li, H.}, \bibinfo{author}{Zhang, X.-Y.}, \bibinfo{author}{Ghosh, B.}, \bibinfo{author}{Hu, C.}, \bibinfo{author}{Wang, Y.-X.}, \bibinfo{author}{Liu, Y.-F.}, \bibinfo{author}{B{\'e}rub{\'e}, D.}, \bibinfo{author}{Dinh, T.}, \bibinfo{author}{Gong, Z.}, \bibinfo{author}{Lien, S.-W.}, \bibinfo{author}{Ho, S.-C.}, \bibinfo{author}{Singh, B.}, \bibinfo{author}{Watanabe, K.}, \bibinfo{author}{Taniguchi, T.}, \bibinfo{author}{Bell, D.~C.}, \bibinfo{author}{Lu, H.-Z.}, \bibinfo{author}{Bansil, A.}, \bibinfo{author}{Lin, H.}, \bibinfo{author}{Chang, T.-R.}, \bibinfo{author}{Zhou, B.~B.}, \bibinfo{author}{Ma, Q.}, \bibinfo{author}{Vishwanath, A.}, \bibinfo{author}{Ni, N.}, and \bibinfo{author}{Xu, S.-Y.} (\bibinfo{year}{2023}). \bibinfo{title}{Axion optical induction of antiferromagnetic order}.
\newblock \bibinfo{journal}{Nature Materials} \emph{\bibinfo{volume}{22}}, \bibinfo{pages}{583--590}. \URLprefix \url{https://doi.org/10.1038/s41563-023-01493-5}. \DOIprefix\doi{10.1038/s41563-023-01493-5}.
\bibitem[{Du et~al.(2023)Du, Xu, Won, Wang, Crooker, Rangan, Bartynski and Cheong}]{Du2023Antiferromagnetic}
\bibinfo{author}{Du, K.}, \bibinfo{author}{Xu, X.}, \bibinfo{author}{Won, C.}, \bibinfo{author}{Wang, K.}, \bibinfo{author}{Crooker, S.~A.}, \bibinfo{author}{Rangan, S.}, \bibinfo{author}{Bartynski, R.}, and \bibinfo{author}{Cheong, S.-W.} (\bibinfo{year}{2023}). \bibinfo{title}{Topological surface magnetism and n{\'e}el vector control in a magnetoelectric antiferromagnet}.
\newblock \bibinfo{journal}{npj Quantum Materials} \emph{\bibinfo{volume}{8}}, \bibinfo{pages}{17}. \URLprefix \url{https://doi.org/10.1038/s41535-023-00551-0}. \DOIprefix\doi{10.1038/s41535-023-00551-0}.
\bibitem[{Satoh et~al.(2010)Satoh, Cho, Iida, Shimura, Kuroda, Ueda, Ueda, Ivanov, Nori and Fiebig}]{satoh2010spin}
\bibinfo{author}{Satoh, T.}, \bibinfo{author}{Cho, S.-J.}, \bibinfo{author}{Iida, R.}, \bibinfo{author}{Shimura, T.}, \bibinfo{author}{Kuroda, K.}, \bibinfo{author}{Ueda, H.}, \bibinfo{author}{Ueda, Y.}, \bibinfo{author}{Ivanov, B.~A.}, \bibinfo{author}{Nori, F.}, and \bibinfo{author}{Fiebig, M.} (\bibinfo{year}{2010}). \bibinfo{title}{{S}pin {O}scillations in {A}ntiferromagnetic $\mathrm{NiO}$ {T}riggered by {C}ircularly {P}olarized {L}ight}.
\newblock \bibinfo{journal}{Phys. Rev. Lett.} \emph{\bibinfo{volume}{105}}, \bibinfo{pages}{077402}. \URLprefix \url{https://link.aps.org/doi/10.1103/PhysRevLett.105.077402}. \DOIprefix\doi{10.1103/PhysRevLett.105.077402}.
\bibitem[{Kampfrath et~al.(2011)Kampfrath, Sell, Klatt, Pashkin, M{\"a}hrlein, Dekorsy, Wolf, Fiebig, Leitenstorfer and Huber}]{kampfrath2011coherent}
\bibinfo{author}{Kampfrath, T.}, \bibinfo{author}{Sell, A.}, \bibinfo{author}{Klatt, G.}, \bibinfo{author}{Pashkin, A.}, \bibinfo{author}{M{\"a}hrlein, S.}, \bibinfo{author}{Dekorsy, T.}, \bibinfo{author}{Wolf, M.}, \bibinfo{author}{Fiebig, M.}, \bibinfo{author}{Leitenstorfer, A.}, and \bibinfo{author}{Huber, R.} (\bibinfo{year}{2011}). \bibinfo{title}{Coherent terahertz control of antiferromagnetic spin waves}.
\newblock \bibinfo{journal}{Nat. Photon.} \emph{\bibinfo{volume}{5}}, \bibinfo{pages}{31--34}. \URLprefix \url{https://doi.org/10.1038/nphoton.2010.259}. \DOIprefix\doi{10.1038/nphoton.2010.259}.
\bibitem[{Zvezdin(1979)}]{zvezdin1979dynamics}
\bibinfo{author}{Zvezdin, A.~K.} (\bibinfo{year}{1979}). \bibinfo{title}{Dynamics of domain walls in weak ferromagnets}.
\newblock \bibinfo{journal}{Pisma Zh. Exp. Teor. Fiz.} \emph{\bibinfo{volume}{29}}, \bibinfo{pages}{605}.
\bibitem[{Zvezdin(2017)}]{zvezdin2017dynamics}
\bibinfo{author}{Zvezdin, A.~K.} (\bibinfo{year}{2017}). \bibinfo{title}{Dynamics of domain walls in weak ferromagnets}.
\newblock \bibinfo{journal}{arXiv preprint arXiv:1703.01502}.
\bibitem[{Mashkovich et~al.(2021)Mashkovich, Grishunin, Dubrovin, Zvezdin, Pisarev and Kimel}]{mashkovich2021terahertz}
\bibinfo{author}{Mashkovich, E.~A.}, \bibinfo{author}{Grishunin, K.~A.}, \bibinfo{author}{Dubrovin, R.~M.}, \bibinfo{author}{Zvezdin, A.~K.}, \bibinfo{author}{Pisarev, R.~V.}, and \bibinfo{author}{Kimel, A.~V.} (\bibinfo{year}{2021}). \bibinfo{title}{Terahertz light--driven coupling of antiferromagnetic spins to lattice}.
\newblock \bibinfo{journal}{Science} \emph{\bibinfo{volume}{374}}, \bibinfo{pages}{1608--1611}. \URLprefix \url{https://doi.org/10.1126/science.abk1121}. \DOIprefix\doi{10.1126/science.abk1121}.
\bibitem[{Behovits et~al.(2023)Behovits, Chekhov, Bodnar, Gueckstock, Reimers, Lytvynenko, Skourski, Wolf, Seifert, Gomonay, Kl{\"a}ui, Jourdan and Kampfrath}]{Behovits2023terahertz}
\bibinfo{author}{Behovits, Y.}, \bibinfo{author}{Chekhov, A.~L.}, \bibinfo{author}{Bodnar, S.~Y.}, \bibinfo{author}{Gueckstock, O.}, \bibinfo{author}{Reimers, S.}, \bibinfo{author}{Lytvynenko, Y.}, \bibinfo{author}{Skourski, Y.}, \bibinfo{author}{Wolf, M.}, \bibinfo{author}{Seifert, T.~S.}, \bibinfo{author}{Gomonay, O.}, \bibinfo{author}{Kl{\"a}ui, M.}, \bibinfo{author}{Jourdan, M.}, and \bibinfo{author}{Kampfrath, T.} (\bibinfo{year}{2023}). \bibinfo{title}{Terahertz n{\'e}el spin-orbit torques drive nonlinear magnon dynamics in antiferromagnetic $\mathrm{Mn}_{2}{Au}$}.
\newblock \bibinfo{journal}{Nature Communications} \emph{\bibinfo{volume}{14}}, \bibinfo{pages}{6038}. \URLprefix \url{https://doi.org/10.1038/s41467-023-41569-z}. \DOIprefix\doi{10.1038/s41467-023-41569-z}.
\bibitem[{Astrov(1960)}]{astrov1960magnetoelectric}
\bibinfo{author}{Astrov, D.~N.} (\bibinfo{year}{1960}). \bibinfo{title}{The magnetoelectric effect in antiferromagnetics}.
\newblock \bibinfo{journal}{Sov. Phys. JETP} \emph{\bibinfo{volume}{11}}, \bibinfo{pages}{708--709}. \URLprefix \url{http://www.jetp.ras.ru/cgi-bin/dn/e_011_03_0708.pdf}.
\bibitem[{Astrov(1961)}]{astrov1961magnetoelectric}
\bibinfo{author}{Astrov, D.~N.} (\bibinfo{year}{1961}). \bibinfo{title}{Magnetoelectric effect in chromium oxide}.
\newblock \bibinfo{journal}{Sov. Phys. JETP} \emph{\bibinfo{volume}{13}}, \bibinfo{pages}{729--733}. \URLprefix \url{http://jetp.ras.ru/cgi-bin/dn/e_013_04_0729.pdf}.
\bibitem[{Foner(1963)}]{foner1963high}
\bibinfo{author}{Foner, S.} (\bibinfo{year}{1963}). \bibinfo{title}{{H}igh-{F}ield {A}ntiferromagnetic {R}esonance in $\mathrm{Cr}_{2}\mathrm{O}_{3}$}.
\newblock \bibinfo{journal}{Phys. Rev.} \emph{\bibinfo{volume}{130}}, \bibinfo{pages}{183--197}. \URLprefix \url{https://link.aps.org/doi/10.1103/PhysRev.130.183}. \DOIprefix\doi{10.1103/PhysRev.130.183}.
\bibitem[{Alikhanov et~al.(1969)Alikhanov, Dimitrijevi{\'c}, Kowalska, Kra{\'s}nicki, R{\.z}any, Todorovi{\'c} and Wanic}]{alikhanov1969neutron}
\bibinfo{author}{Alikhanov, R.}, \bibinfo{author}{Dimitrijevi{\'c}, {\v{Z}}.}, \bibinfo{author}{Kowalska, A.}, \bibinfo{author}{Kra{\'s}nicki, S.}, \bibinfo{author}{R{\.z}any, H.}, \bibinfo{author}{Todorovi{\'c}, J.}, and \bibinfo{author}{Wanic, A.} (\bibinfo{year}{1969}). \bibinfo{title}{{N}eutron {I}nvestigation of the {S}pin {S}ystem {D}ynamics in $\alpha$-$\mathrm{Cr}_{2}\mathrm{O}_{3}$}.
\newblock \bibinfo{journal}{physica status solidi (b)} \emph{\bibinfo{volume}{32}}, \bibinfo{pages}{41--48}. \URLprefix \url{https://doi.org/10.1002/pssb.19690320105}. \DOIprefix\doi{10.1002/pssb.19690320105}.
\bibitem[{Samuelsen et~al.(1970)Samuelsen, Hutchings and Shirane}]{samuelsen1970inelastic}
\bibinfo{author}{Samuelsen, E.~J.}, \bibinfo{author}{Hutchings, M.~T.}, and \bibinfo{author}{Shirane, G.} (\bibinfo{year}{1970}). \bibinfo{title}{Inelastic neutron scattering investigation of spin waves and magnetic interactions in $\mathrm{Cr}_{2}\mathrm{O}_{3}$}.
\newblock \bibinfo{journal}{Physica} \emph{\bibinfo{volume}{48}}, \bibinfo{pages}{13--42}. \URLprefix \url{https://doi.org/10.1016/0031-8914(70)90158-8}. \DOIprefix\doi{10.1016/0031-8914(70)90158-8}.
\bibitem[{Biao et~al.(2023)Biao, Jun, Yuan-zhe, Di and Qi}]{biao2023low}
\bibinfo{author}{Biao, D.}, \bibinfo{author}{Jun, C.}, \bibinfo{author}{Yuan-zhe, T.}, \bibinfo{author}{Di, W.}, and \bibinfo{author}{Qi, Z.} (\bibinfo{year}{2023}). \bibinfo{title}{{L}ow-{F}requency {R}aman {D}etection of {A}ntiferromagnetic {S}pin {W}aves in $\mathrm{Cr}_{2}\mathrm{O}_{3}$}.
\newblock \bibinfo{journal}{Prog. Phys.} \emph{\bibinfo{volume}{43}}, \bibinfo{pages}{142--150}. \URLprefix \url{https://doi.org/10.13725/j.cnki.pip.2023.05.002}. \DOIprefix\doi{10.13725/j.cnki.pip.2023.05.002}.
\bibitem[{Sun et~al.(2024)Sun, Wei, Su, Abernathy and Li}]{qiyang2024temperature}
\bibinfo{author}{Sun, Q.}, \bibinfo{author}{Wei, B.}, \bibinfo{author}{Su, Y.}, \bibinfo{author}{Abernathy, D.~L.}, and \bibinfo{author}{Li, C.} (\bibinfo{year}{2024}). \bibinfo{title}{Temperature-dependent spin dynamics in $\mathrm{Cr}_{2}\mathrm{O}_{3}$}.
\newblock \bibinfo{journal}{Phys. Rev. B} \emph{\bibinfo{volume}{110}}, \bibinfo{pages}{064413}. \URLprefix \url{https://link.aps.org/doi/10.1103/PhysRevB.110.064413}. \DOIprefix\doi{10.1103/PhysRevB.110.064413}.
\bibitem[{Baierl et~al.(2016)Baierl, Hohenleutner, Kampfrath, Zvezdin, Kimel, Huber and Mikhaylovskiy}]{baierl2016nonlinear}
\bibinfo{author}{Baierl, S.}, \bibinfo{author}{Hohenleutner, M.}, \bibinfo{author}{Kampfrath, T.}, \bibinfo{author}{Zvezdin, A.~K.}, \bibinfo{author}{Kimel, A.~V.}, \bibinfo{author}{Huber, R.}, and \bibinfo{author}{Mikhaylovskiy, R.~V.} (\bibinfo{year}{2016}). \bibinfo{title}{Nonlinear spin control by terahertz-driven anisotropy fields}.
\newblock \bibinfo{journal}{Nat. Photon.} \emph{\bibinfo{volume}{10}}, \bibinfo{pages}{715--718}. \URLprefix \url{https://doi.org/10.1038/nphoton.2016.181}. \DOIprefix\doi{10.1038/nphoton.2016.181}.
\bibitem[{Belov et~al.(1993)Belov, Vorob'ev, Kadomtseva, Popov and Zvezdin}]{belov1993magnetoelectric}
\bibinfo{author}{Belov, D.~V.}, \bibinfo{author}{Vorob'ev, G.~P.}, \bibinfo{author}{Kadomtseva, A.~M.}, \bibinfo{author}{Popov, Y.~F.}, and \bibinfo{author}{Zvezdin, A.~K.} (\bibinfo{year}{1993}). \bibinfo{title}{Magnetoelectric effect in the spin-flop phase of $\mathrm{Cr}_{2}\mathrm{O}_{3}$ and the problem of determining the magnetic structure}.
\newblock \bibinfo{journal}{JETP Lett.} \emph{\bibinfo{volume}{58}}. \URLprefix \url{http://jetpletters.ru/ps/1191/article_17975.pdf}.
\bibitem[{Volger(1952)}]{volger1952anomalous}
\bibinfo{author}{Volger, J.} (\bibinfo{year}{1952}). \bibinfo{title}{{A}nomalous {S}pecific {H}eat of {C}hromium {O}xide ($\mathrm{Cr}_{2}\mathrm{O}_{3}$) at the {A}ntiferromagnetic {C}urie {T}emperature}.
\newblock \bibinfo{journal}{Nature} \emph{\bibinfo{volume}{170}}, \bibinfo{pages}{1027--1027}. \URLprefix \url{https://doi.org/10.1038/1701027a0}. \DOIprefix\doi{10.1038/1701027a0}.
\bibitem[{{McGuire} et~al.(1956){McGuire}, Scott and Grannis}]{mcguire1956antiferromagnetism}
\bibinfo{author}{{McGuire}, T.~R.}, \bibinfo{author}{Scott, E.~J.}, and \bibinfo{author}{Grannis, F.~H.} (\bibinfo{year}{1956}). \bibinfo{title}{{A}ntiferromagnetism in a $\mathrm{Cr}_{2}\mathrm{O}_{3}$ {C}rystal}.
\newblock \bibinfo{journal}{Phys. Rev.} \emph{\bibinfo{volume}{102}}, \bibinfo{pages}{1000--1003}. \URLprefix \url{https://link.aps.org/doi/10.1103/PhysRev.102.1000}. \DOIprefix\doi{10.1103/PhysRev.102.1000}.
\bibitem[{Brockhouse(1953)}]{brockhouse1953antiferromagnetic}
\bibinfo{author}{Brockhouse, B.~N.} (\bibinfo{year}{1953}). \bibinfo{title}{Antiferromagnetic structure in $\mathrm{Cr}_{2}\mathrm{O}_{3}$}.
\newblock \bibinfo{journal}{J. Chem. Phys.} \emph{\bibinfo{volume}{21}}, \bibinfo{pages}{961--962}. \URLprefix \url{https://doi.org/10.1063/1.1699098}. \DOIprefix\doi{10.1063/1.1699098}.
\bibitem[{Corliss et~al.(1965)Corliss, Hastings, Nathans and Shirane}]{corliss1965magnetic}
\bibinfo{author}{Corliss, L.~M.}, \bibinfo{author}{Hastings, J.~M.}, \bibinfo{author}{Nathans, R.}, and \bibinfo{author}{Shirane, G.} (\bibinfo{year}{1965}). \bibinfo{title}{Magnetic structure of $\mathrm{Cr}_{2}\mathrm{O}_{3}$}.
\newblock \bibinfo{journal}{J. Appl. Phys.} \emph{\bibinfo{volume}{36}}, \bibinfo{pages}{1099--1100}. \URLprefix \url{https://doi.org/10.1063/1.1714118}. \DOIprefix\doi{10.1063/1.1714118}.
\bibitem[{Bousquet et~al.(2024)Bousquet, Leli{\`e}vre-Berna, Qureshi, Soh, Spaldin, Urru, Verbeek and Weber}]{bousquet2024sign}
\bibinfo{author}{Bousquet, E.}, \bibinfo{author}{Leli{\`e}vre-Berna, E.}, \bibinfo{author}{Qureshi, N.}, \bibinfo{author}{Soh, J.-R.}, \bibinfo{author}{Spaldin, N.~A.}, \bibinfo{author}{Urru, A.}, \bibinfo{author}{Verbeek, X.~H.}, and \bibinfo{author}{Weber, S.~F.} (\bibinfo{year}{2024}). \bibinfo{title}{On the sign of the linear magnetoelectric coefficient in $\mathrm{Cr}_{2}\mathrm{O}_{3}$}.
\newblock \bibinfo{journal}{J. Phys. Condens. Matter} \emph{\bibinfo{volume}{36}}, \bibinfo{pages}{155701}. \URLprefix \url{https://doi.org/10.1088/1361-648X/ad1a59}. \DOIprefix\doi{10.1088/1361-648X/ad1a59}.
\bibitem[{Dzyaloshinskii(1960)}]{dzyaloshinskii1960magneto}
\bibinfo{author}{Dzyaloshinskii, I.~E.} (\bibinfo{year}{1960}). \bibinfo{title}{On the magneto-electrical effects in antiferromagnets}.
\newblock \bibinfo{journal}{Sov. Phys. JETP} \emph{\bibinfo{volume}{10}}, \bibinfo{pages}{628--629}. \DOIprefix\doi{http://jetp.ras.ru/cgi-bin/dn/e_010_03_0628.pdf}.
\bibitem[{Fiebig et~al.(1995)Fiebig, Fr{\"o}hlich, Sluyterman~v. L and Pisarev}]{fiebig1995domain}
\bibinfo{author}{Fiebig, M.}, \bibinfo{author}{Fr{\"o}hlich, D.}, \bibinfo{author}{Sluyterman~v. L, G.}, and \bibinfo{author}{Pisarev, R.~V.} (\bibinfo{year}{1995}). \bibinfo{title}{Domain topography of antiferromagnetic $\mathrm{Cr}_{2}\mathrm{O}_{3}$ by second-harmonic generation}.
\newblock \bibinfo{journal}{Appl. Phys. Lett.} \emph{\bibinfo{volume}{66}}, \bibinfo{pages}{2906--2908}. \URLprefix \url{https://doi.org/10.1063/1.113699}. \DOIprefix\doi{10.1063/1.113699}.
\bibitem[{Turov et~al.(2001)Turov, Kolchanov, Menshenin, Mirsaev and Nikolaev}]{turov2001symmetry}
\bibinfo{author}{Turov, E.~A.}, \bibinfo{author}{Kolchanov, A.~V.}, \bibinfo{author}{Menshenin, V.~V.}, \bibinfo{author}{Mirsaev, I.~F.}, and \bibinfo{author}{Nikolaev, V.~V.} (\bibinfo{year}{2001}). \bibinfo{title}{{S}ymmetry and {P}hysical {P}roperties of {A}ntiferromagnets}.
\newblock \bibinfo{journal}{Fizmatlit, Moscow}.
\bibitem[{Fradkin(2013)}]{fradkin2013field}
\bibinfo{author}{Fradkin, E.}
\newblock \bibinfo{title}{{F}ield {T}heories of {C}ondensed {M}atter {P}hysics}.
\newblock \bibinfo{publisher}{Cambridge University Press} (\bibinfo{year}{2013}).
\bibitem[{Zvezdin et~al.(2024)Zvezdin, Dubrovin and Kimel}]{zvezdin2024giant}
\bibinfo{author}{Zvezdin, A.~K.}, \bibinfo{author}{Dubrovin, R.~M.}, and \bibinfo{author}{Kimel, A.~V.} (\bibinfo{year}{2024}). \bibinfo{title}{{G}iant {P}arametric {A}mplification of the {I}nverse {C}otton--{M}outon {E}ffect in {A}ntiferromagnetic {C}rystals}.
\newblock \bibinfo{journal}{JETP Lett.} \emph{\bibinfo{volume}{119}}, \bibinfo{pages}{363--371}. \URLprefix \url{https://doi.org/10.1134/S0021364023604050}. \DOIprefix\doi{10.1134/S0021364023604050}.
\bibitem[{Mukhin et~al.(1997)Mukhin, Travkin and Lebedev}]{mukhin1997bwo}
\bibinfo{author}{Mukhin, A.~A.}, \bibinfo{author}{Travkin, V.~D.}, and \bibinfo{author}{Lebedev, S.~P.} (\bibinfo{year}{1997}). \bibinfo{title}{{BWO} quasi-optical spectroscopy of $\mathrm{Cr}_{2}\mathrm{O}_{3}$: {G}yrotropic birefringence at antiferromagnetic resonance}.
\newblock \bibinfo{journal}{Ferroelectrics} \emph{\bibinfo{volume}{204}}, \bibinfo{pages}{261--267}. \DOIprefix\doi{10.1080/00150199708222206}.
\bibitem[{Birss et~al.(1964)}]{birss1964symmetry}
\bibinfo{author}{Birss, R.~R.} et~al.
\newblock \bibinfo{title}{Symmetry and magnetism} vol. \bibinfo{volume}{863}.
\newblock \bibinfo{publisher}{North-Holland Amsterdam} (\bibinfo{year}{1964}).

\end{thebibliography}
\end{document}